\documentclass{ws-ijmpe}

\begin{document}

\markboth{Takao Sakaguchi}{System size and energy dependence of high $p_T$ hadron production}

\catchline{}{}{}{}{}

\title{System size and energy dependence of high $p_T$ hadron production measured with PHENIX experiment at RHIC}

\author{\footnotesize Takao Sakaguchi, for the PHENIX Collaboration}

\address{Physics Department, Brookhaven National Laboratory,\\ Upton, NY 11973, U.S.A.}

\maketitle

\begin{history}
\received{(received date)}
\revised{(revised date)}
\end{history}

\begin{abstract}
PHENIX has measured high transverse momentum ($p_T$) identified hadrons
in different collision species and energies in the last
five RHIC runs. The systematic study of the high $p_T$ hadron
production provides an idea on interaction of hard scattered partons
and the matter created in relativistic heavy ion collision.
The $\eta/\pi^{0}$ ratio is measured in Au+Au collisions, which
 gives a hint on the system thermalization and particle production.
A future measurement of hadron and photon measurement is discussed.
\end{abstract}

\section{Introduction}
One of the essential questions in relativistic heavy ion collisions
is how the system is thermalized from the early stages of collisions.
A recent theoretical implication of direct photon data
from the PHENIX experiment~\cite{ref1,ref2} says that the thermalization
time of the partonic system is $\tau\sim$0.26\,fm/$c$~\cite{ref3}.
It would imply that the partonic system is almost suddenly thermalized
after initial collisions.
This rapid thermalization is also suggested from a quark and kinetic
energy scaling of strong elliptic flow measured by the PHENIX
experiment~\cite{ref4}. The collective behavior of partons may be
realized already at an early stage when the collisional area
is most anisotropy defined by the initial collision geometry.

One idea to investigate the state of the matter created is to measure the
interaction of well-calibrated probe, such as hard scattered partons, with
the matter. In high energy p+p collisions, high transverse momentum ($p_{T}$)
hadrons are mostly produced from jets, the fragments of hard scattered partons.
In the presence of a hot dense matter produced in relativistic heavy ion
collisions, such partons may lose energy, and therefore the fragmentation
function is modified. It may result in a different particle yield
ratio as a function of momentum compared to p+p collisions. Hard scattered
partons may also combine with partons in the medium, and produce mid-$p_T$
(2-5\,GeV/$c$) particles.  It is a different mechanism from the one expected
in p+p collisions and can be said as an interaction of the partons with
the medium as well~\cite{ref5}.

A systematic study of identified hadron production at high $p_T$ provides
many intriguing information on the matter. In this presentation, the recent
measurement on high $p_T$ identified hadrons at a mid-rapidity is reviewed
and discussed. A possible future measurement is also presented.

\section{Measurement of high $p_T$ $\pi^0$ production in Au+Au and Cu+Cu collisions}
\begin{figure}[h]
\begin{center}
\begin{minipage}{55mm}
\centerline{\psfig{file=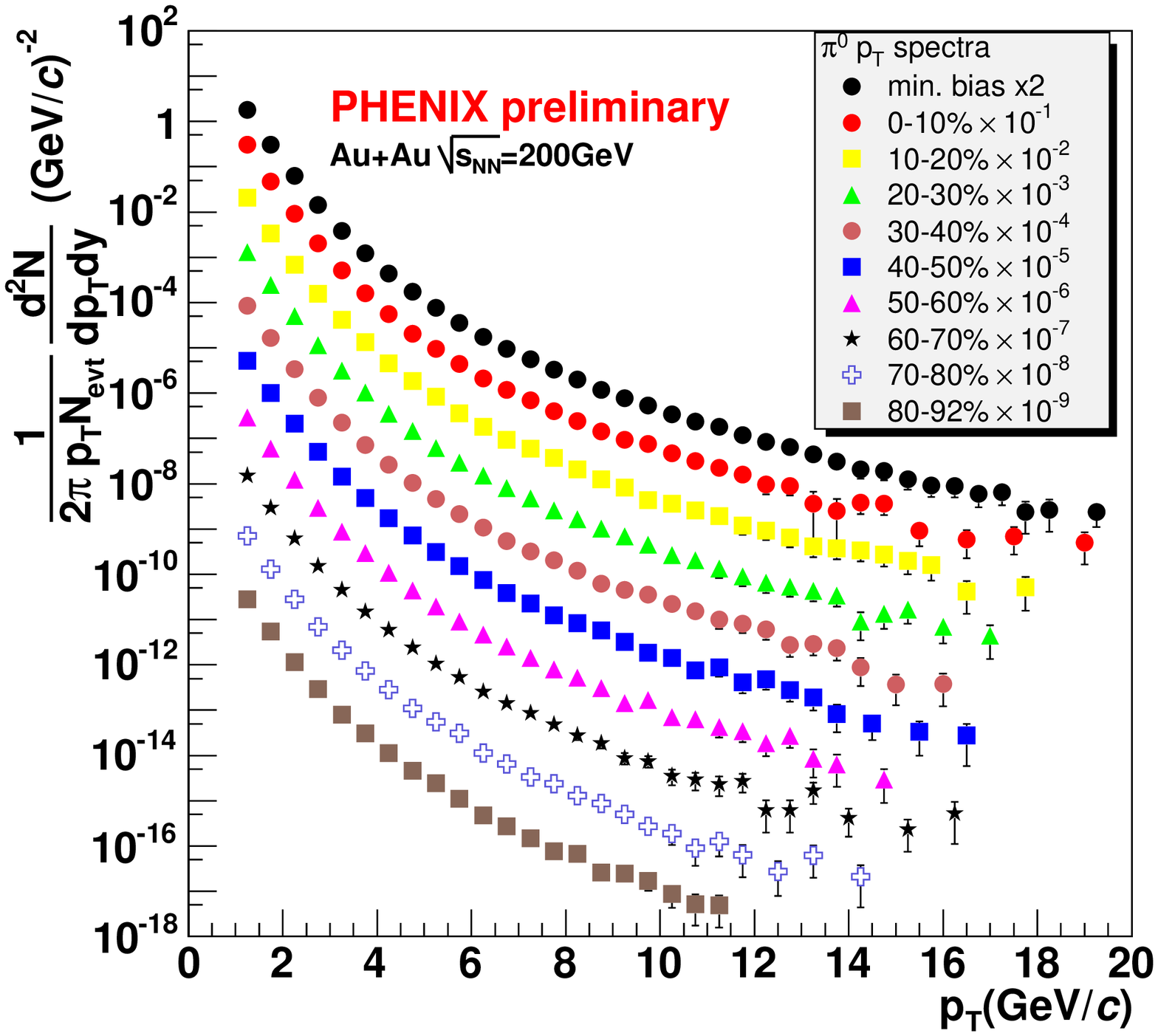,width=6.5cm}}
\caption{$\pi^{0}$ $p_T$ spectra in Au+Au collisions at $\sqrt{s_{NN}}$=200\,GeV.}
\label{fig1}
\end{minipage}
\hspace{5mm}
\begin{minipage}{55mm}
\centerline{\psfig{file=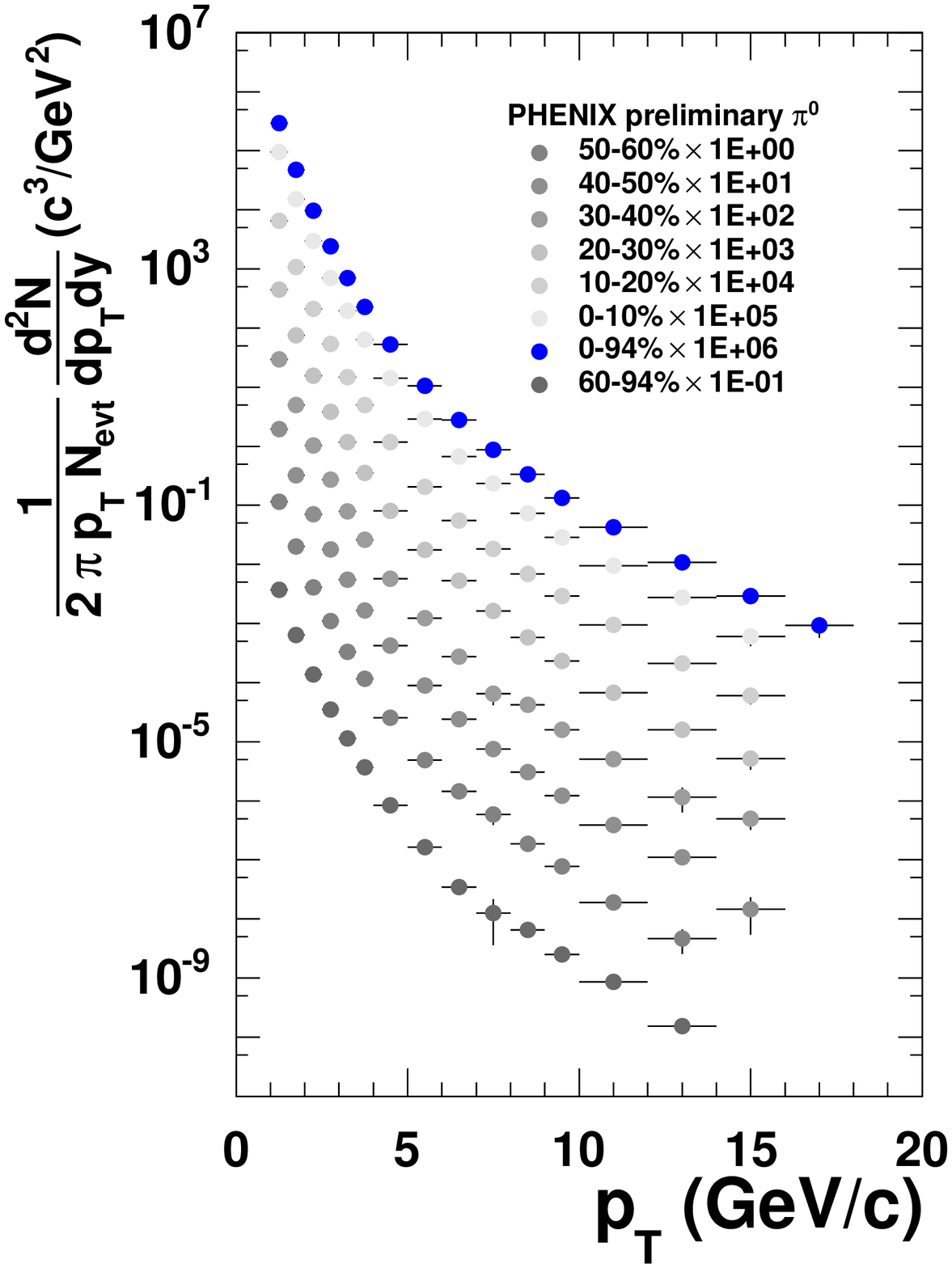,width=5cm}}
\caption{$\pi^{0}$ $p_T$ spectra in Cu+Cu collisions at $\sqrt{s_{NN}}$=200\,GeV.}
\label{fig2}
\end{minipage}
\end{center}
\end{figure}

PHENIX has measured $p_T$ spectra of $\pi^0$ at different energies and
different collision systems. Figure~\ref{fig1} shows the $p_T$ spectra of
$\pi^0$ in Au+Au collisions at $\sqrt{s_{NN}}$=200\,GeV for different
centralities from RHIC Year-4 run. The measurement is extended
up to 20\,GeV/$c$, compared to year-2 run, with 900\,M minimum bias
collisions. Figure~\ref{fig2} shows the $p_T$ spectra of $\pi^0$
in Cu+Cu collisions at the same c.m.s. energy from RHIC Year-5 run.
These measurement were made through $\pi^{0} \rightarrow \gamma\gamma$ channel.

The change of the spectra shape as a function of centrality can quantitatively
be evaluated using a nuclear modification factor ($R_{AA}$) defined as:
\[R_{AA} = \frac{(1/N_{evt}) dN^{AuAu}/dydp_{T}}{<N_{coll}> (1/\sigma^{pp}_{inel}) d\sigma/dydp_{T}} \]
where, $<N_{coll}>$ is the average number of binary nucleon-nucleon collisions.
The ratio is expected to be unity if particles are produced just by
hard scattering, and no nuclear effect is involved.

\begin{figure}[th]
\begin{center}
\begin{minipage}{55mm}
\centerline{\psfig{file=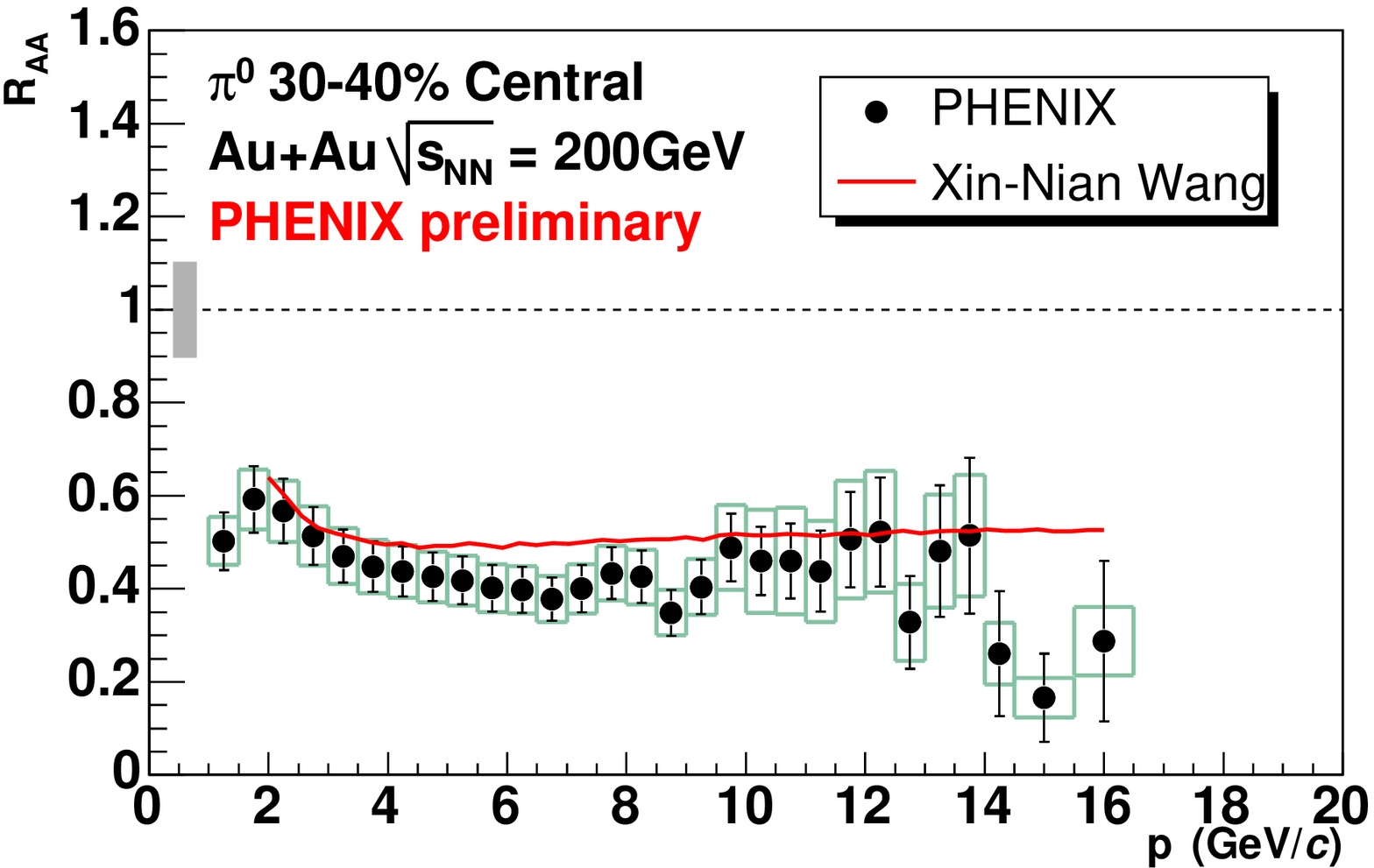,width=7cm}}
\caption{Nuclear modification factor ($R_{AA}$) for $\pi^{0}$ for 30-40\,\% centrality in Au+Au collisions at $\sqrt{s_{NN}}$=200\,GeV.}
\label{fig3}
\end{minipage}
\hspace{5mm}
\begin{minipage}{55mm}
\centerline{\psfig{file=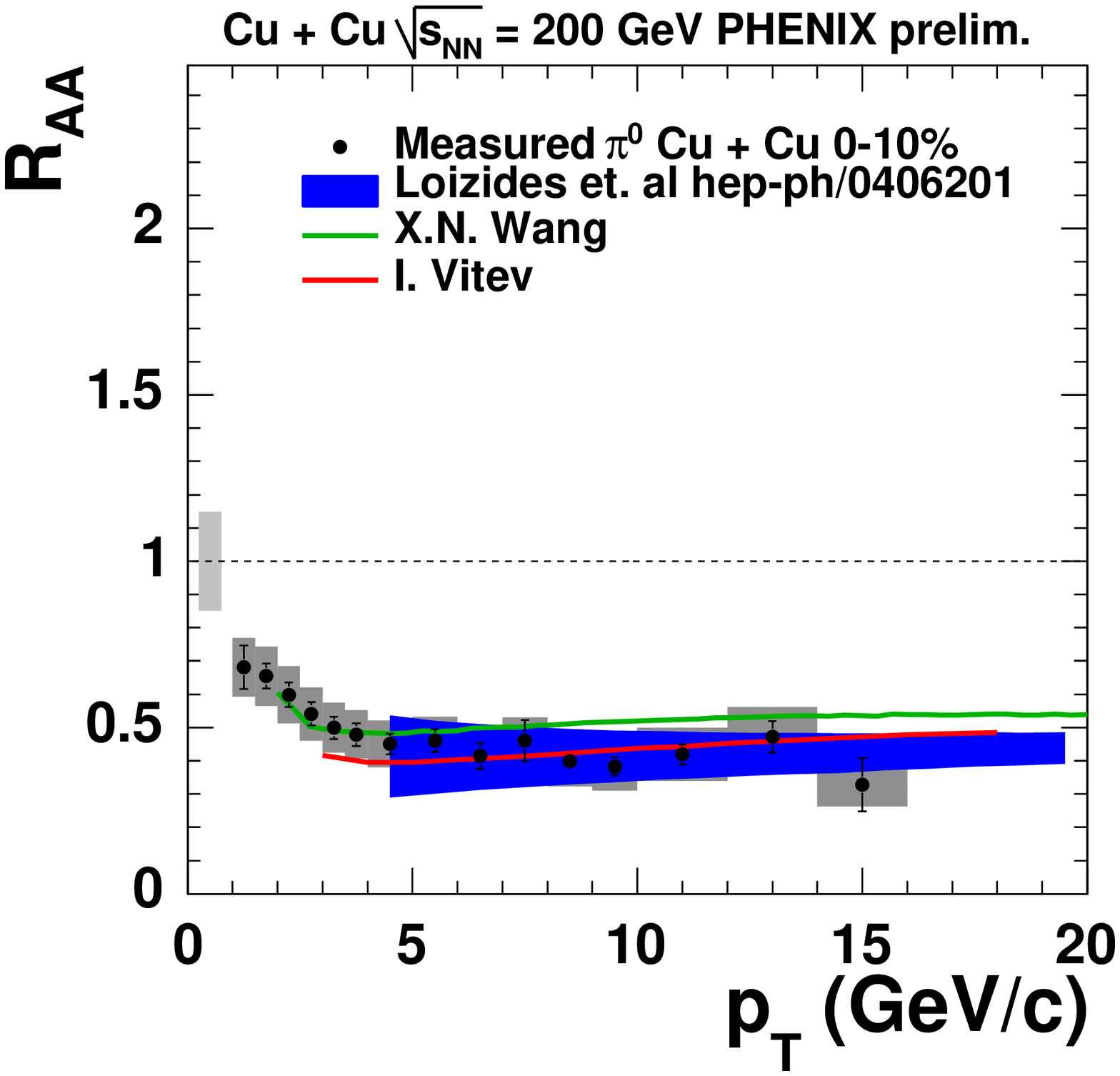,width=5cm}}
\vspace{-2mm}
\caption{Nuclear modification factor ($R_{AA}$) for $\pi^{0}$ for 0-10\,\% centrality in Cu+Cu collisions at $\sqrt{s_{NN}}$=200\,GeV.}
\label{fig4}
\end{minipage}
\end{center}
\end{figure}

\begin{figure}[th]
\begin{center}
\begin{minipage}{55mm}
\centerline{\psfig{file=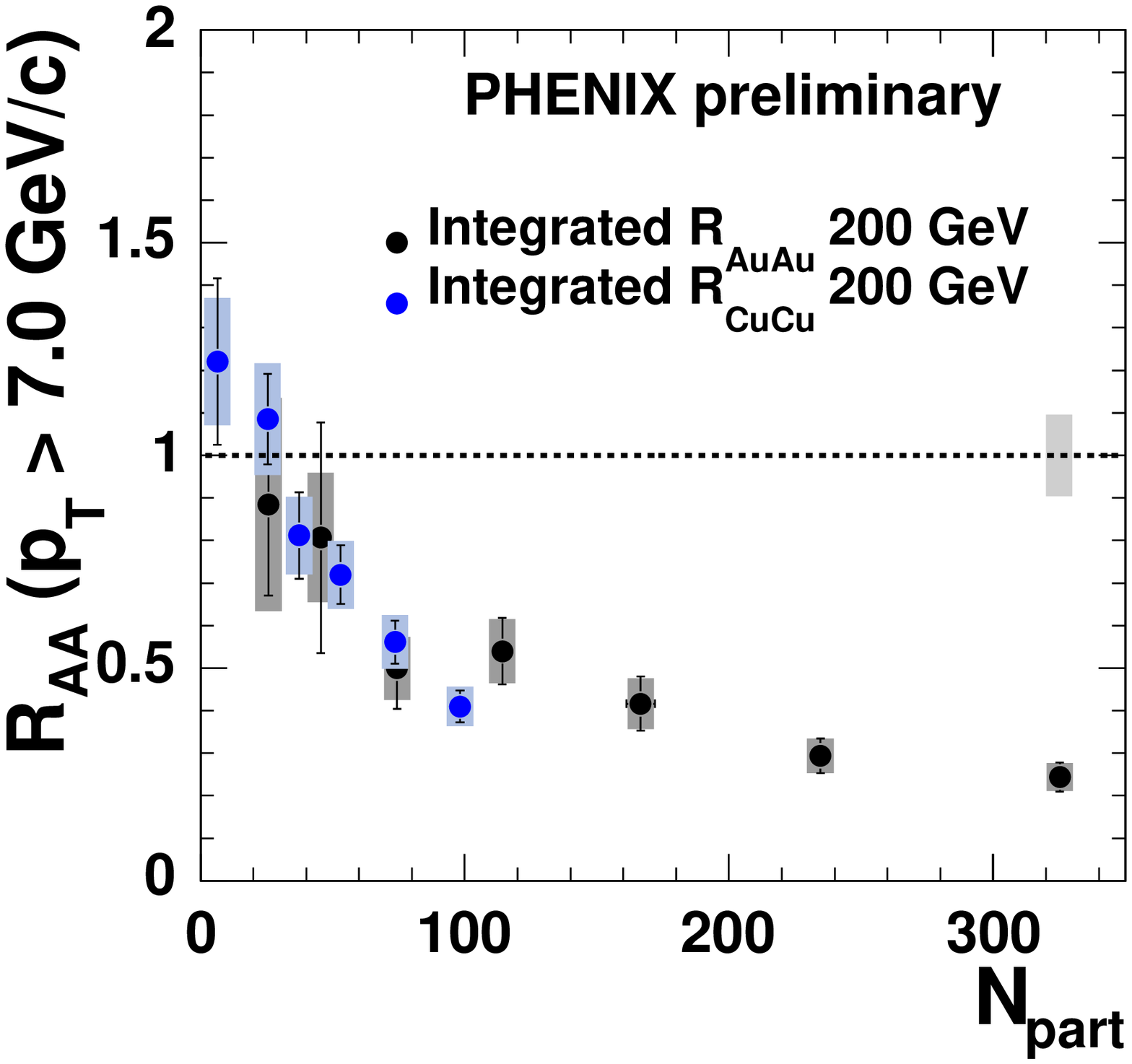,width=6cm}}
\vspace{-3mm}
\caption{Integrated $R_{AA}$ $\pi^{0}$ in Au+Au and Cu+Cu collisions for $p_T>$7\,GeV/$c$ at $\sqrt{s_{NN}}$=200\,GeV.}
\label{fig5}
\end{minipage}
\hspace{5mm}
\begin{minipage}{55mm}
\centerline{\psfig{file=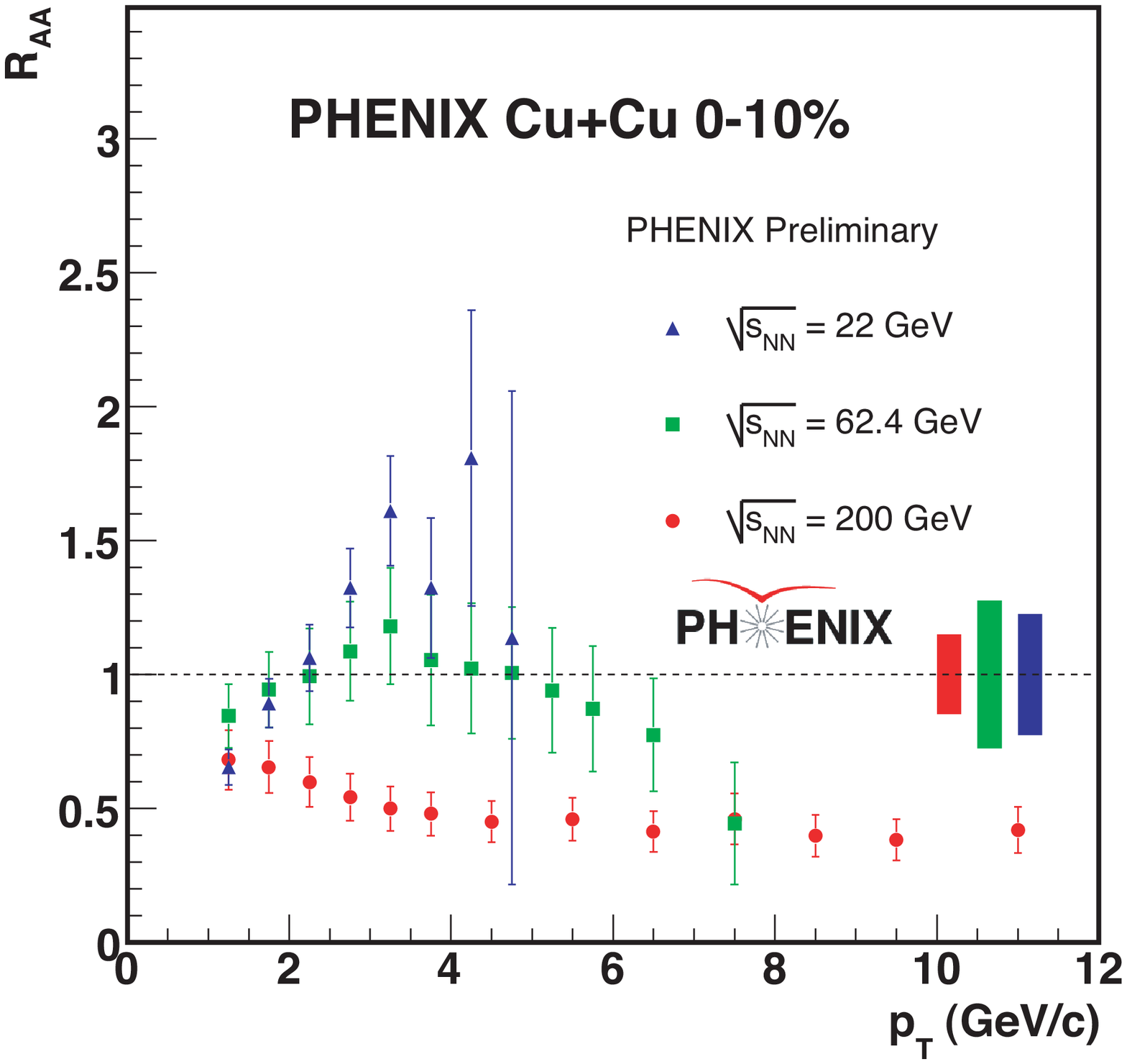,width=5.5cm}}
\caption{$R_{AA}$ for $\pi^{0}$ in Cu+Cu collisions at $\sqrt{s_{NN}}$=22, 62.4 and 200\,GeV.}
\label{fig6}
\end{minipage}
\end{center}
\end{figure}

Figures~\ref{fig3} and ~\ref{fig4} show $R_{AA}$ for
$\pi^{0}$ in Au+Au and Cu+Cu collisions at $\sqrt{s_{NN}}=200$\,GeV.
The centrality in two plots are different (30-40\,\% for Au+Au and 0-10\,\%
for Cu+Cu), but the number of participant nucleons are around same.
It is seen that the suppression factors in the two systems are almost same.
It is even clearly seen in the integrated $R_{AA}$ for $p_T>$7\,GeV/$c$ as
shown in Fig.~\ref{fig5}. At a high $p_T$ ($p_T$$>$$\sim$5\,GeV/$c$),
$\pi^{0}$'s are mostly from jets, and therefore should essentially have
the same slopes expected from a hard scattering, over centralities.
The same $R_{AA}$ at the same number of participants would imply
that the amount of the matter that hard scattered partons suffers
in both systems are same even though the shape of collisional regions
are different. Comparisons with some
model calculations are also overlaid in the plots~\cite{ref6,ref7,ref8}.

The energy dependence of hadron suppression has been measured in Cu+Cu
collisions (Fig.~\ref{fig6}). It turned out that the suppression is degraded
as energy decreases, and the Cronin-like enhancement becomes visible.
It suggests the energy loss is smaller in the lower c.m.s. energy,
but it is also possible that the particle production at a lower energy
does not completely follow the number of binary collision scaling in heavy
ion collisions. The direct photon results confirmed that the probability
of an initial hard scattering that produces high $p_T$ hadrons is scaled as
the number of binary collisions only at $\sqrt{s_{NN}}$=200\,GeV. We should
measure direct photons at lower energy as well to make a conclusive statement.

\section{$\eta$ spectra and $\eta/\pi^{0}$ ratio in Au+Au collisions}
\begin{figure}[th]
\begin{center}
\centerline{\psfig{file=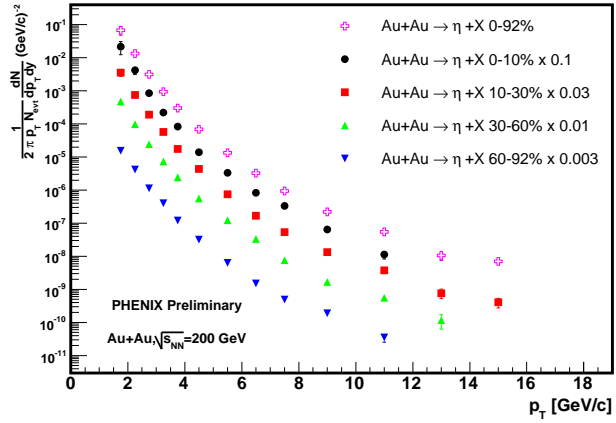,width=9cm}}
\caption{$p_T$ spectra of $\eta$ in Cu+Cu collisions at $\sqrt{s_{NN}}$=200\,GeV.}
\label{fig7}
\end{center}
\end{figure}

\begin{figure}[th]
\begin{center}
\centerline{\psfig{file=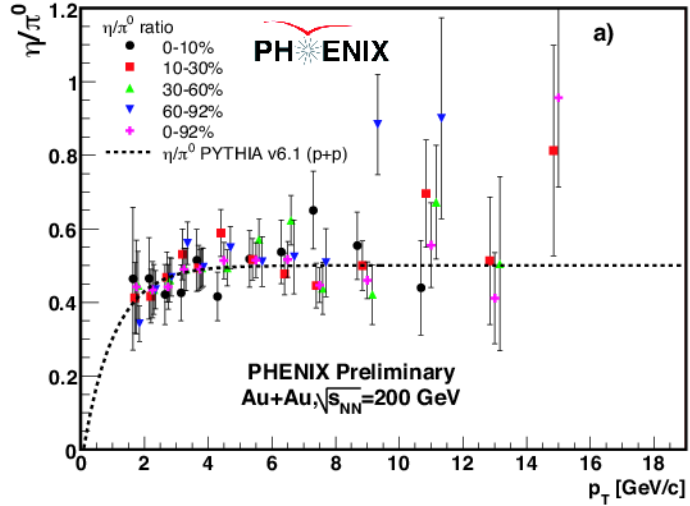,width=9cm}}
\caption{$\eta/\pi^{0}$ ratio in Au+Au collisions at $\sqrt{s_{NN}}$=200\,GeV collisions.}
\label{fig8}
\end{center}
\end{figure}

PHENIX has measured $\eta$ in a wide $p_T$ range of from 1.5\,GeV/$c$
to 15\,GeV/$c$ via $\eta \rightarrow \gamma\gamma$ channel.
Figure~\ref{fig7} shows the $\eta$ spectra over centralites in Au+Au
collisions at $\sqrt{s_{NN}}$=200\,GeV.
The $\eta/\pi^0$ ratio has been particularly of interest for a direct
photon measurement. The value has been used to estimate the yield of
$\eta$ from $\pi^{0}$ yield in case that they are not clearly observed.
Surprisingly, the ratio in Au+Au is well described by PYTHIA~\cite{ref12}
(Fig.~\ref{fig8}), though copious "soft" $\eta$'s is expected
in Au+Au collisions. This implies that either the thermal production of
$\eta$ and $\pi^{0}$ is similar to the one expected from a hard scattering
followed by fragmentation function, or the system is almost thermalized
immediately after the collisions.

\section{Future measurement of the particle production in PHENIX}
One way to investigate the thermalization scenario is to measure the hadron
spectra as a function of rapidity. A theoretical model on the rapid
thermalization assumes an interaction of a nucleus with the strong color
field created by the other incoming nucleus~\cite{ref9}.
The model suggests that the amount of color potential receiving from the
incoming nucleus depends on the rapidity, and thus predicts different
temperature of hadrons at different rapidities. There is a prediction
also for photon production~\cite{ref10}. The prediction says that
the rapidity dependence of photon temperature is different depending on
the system expansion scenarios.

\begin{figure}[th]
\begin{center}
\centerline{\psfig{file=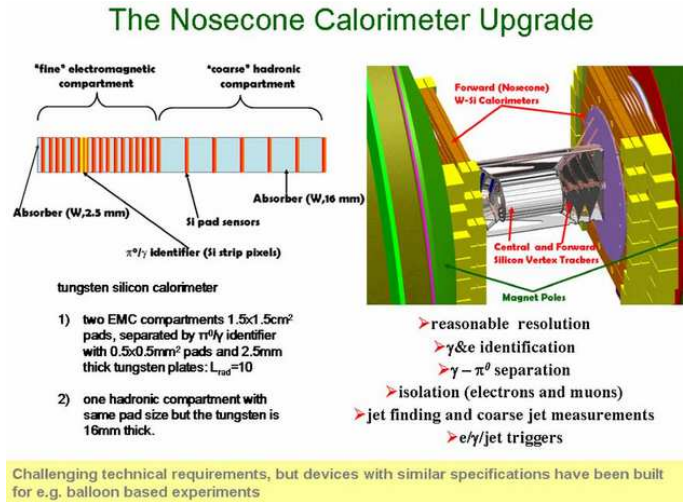,width=9cm}}
\caption{PHENIX nose cone calorimeter.}
\label{fig9}
\end{center}
\end{figure}

The PHENIX experiment has an upgrade project to strengthen the
PID and calorimetric capability at forward rapidity region.
Figure~\ref{fig9} shows a calorimeter that is going to be installed 
into the rapidity of y=1-2 in 2011.
The measurement at forward rapidity region would give hints on
understanding the particle production mechanism at RHIC.

\section{Summary}
PHENIX has measured high transverse momentum ($p_T$) identified hadrons
in different collision species and energies in the last
five RHIC runs. The systematic study of the high $p_T$ hadron
production provides an idea on interaction of hard scattered partons
and the matter created in relativistic heavy ion collision.
The $\eta/\pi^{0}$ ratio is measured in Au+Au collisions, which
 gives a hint on the system thermalization and particle production.

\end{document}